# Single-Laser 32.5 Tbit/s Nyquist WDM Transmission

David Hillerkuss, Rene Schmogrow, Matthias Meyer, Stefan Wolf, Meinert Jordan,
Philipp Kleinow, Nicole Lindenmann, Philipp C. Schindler, Argishti Melikyan,
Xin Yang, Shalva Ben-Ezra, Bend Nebendahl, Michael Dreschmann, Joachim Meyer,
Francesca Parmigiani, Periklis Petropoulos, Bojan Resan, Andreas Oehler,
Kurt Weingarten, Lars Altenhain, Tobias Ellermeyer, Michael Moeller,
Michael Huebner, Juergen Becker, Christian Koos, Wolfgang Freude, and
Juerg Leuthold

*Abstract*—We demonstrate single laser 32.5 Tbit/s 16QAM Nyquist WDM transmission over a total length of 227 km of SMF-28 without optical dispersion compensation. A number of 325 optical carriers are derived from a single laser and encoded with dual-polarization 16QAM data using sinc-shaped Nyquist pulses. As we use no guard bands, the carriers have a spacing of 12.5 GHz equal to the symbol rate or Nyquist bandwidth of the data. We achieve a net spectral efficiency of 6.4 bit/s/Hz using a software-defined transmitter, which generates the electric drive-signals for the electro-optic modulator in real-time.

*Index Terms*—Communication systems; Optical fiber communication; Pulse shaping methods; Quadrature amplitude modulation;

D. Hillerkuss, R. Schmogrow, M. Meyer, S. Wolf, M. Jordan, P. Kleinow, N. Lindenmann, P. C. Schindler, A. Melikyan, M. Dreschmann, J. Meyer, J. Becker, C. Koos, W. Freude, and J. Leuthold are with the Karlsruhe Institute of Technology (KIT), Karlsruhe, Germany (david.hillerkuss@kit.edu).

X. Yang, F. Parmigiani, and P. Petropoulos are with the Optoelectronics Research Centre, University of Southampton, Southampton, United Kingdom.

S. Ben-Ezra is with Finisar Corporation, Nes Ziona, Israel.

B. Nebendahl is with Agilent Technologies, Boeblingen, Germany.

B. Resan, A. Oehler, and K. Weingarten are with Time-Bandwidth Products, Zurich, Switzerland.

L. Altenhain is with the Micram Microelectronic GmbH, Bochum, Germany

T. Ellermeyer was with the Micram Microelectronic GmbH, Bochum, Germany. He is now with Fachhochschule Südwestfalen / University of Applied Sciences, Iserlohn, Germany

M. Moeller is with the Department of Electronics and Circuits, Saarland University, Saarbruecken, Germany

M. Huebner was with Karlsruhe Institute of Technology (KIT), Karlsruhe, Germany. He is now with Ruhr-Universität Bochum, Bochum, Germany

## I. Introduction

Super-channels for multi-Tbit/s transmission are envisioned to play an important role in future optical networks [1]. Such channels typically consist of one carrier or several frequency-locked carriers onto which data are encoded [2-5]. As cost and power consumption are important issues [6-8], a reduced component count is desirable. Therefore, single-laser Tbit/s transmission systems are of special interest.

Up to the year 2009, single-laser Tbit/s systems were mostly implemented using optical time division multiplexing (OTDM) [2, 3]. With OTDM, data rates of up to 10.2 Tbit/s were obtained within an optical bandwidth of roughly 30 nm or 3.75 THz. This corresponds to a net spectral efficiency of 2.6 bit/s/Hz [3]. With TDM, transmission at 10.2 Tbit/s over 29 km of dispersion-managed fiber has been demonstrated.

Since 2005, multicarrier transmission has attracted increasing interest as it offers highest spectral efficiency. In particular, coherent wavelength division multiplexing (CO-WDM) [9] and orthogonal frequency division multiplexing (OFDM) [10-12] have been proposed. The first demonstration of an OFDM-signal beyond 1.0 Tbit/s in 2009 showed a transmission distance of 600 km using standard single mode fiber [13]. The spectral efficiency was 3.3 bit/s/Hz. In 2010, using an optical FFT scheme [14], we were able to encode and detect a 10.8 Tbit/s super-channel [15]. Subsequently, we generated and transmitted an OFDM super-channel with a line rate of 26 Tbit/s over a distance of 50 km of standard single mode fiber with standard dispersion compensating modules [4]. The net spectral efficiency could be increased to 5.0 bit/s/Hz.

Nyquist pulse shaping [16] is an alternate method to improve the transmission performance. Such a pulse shaping can increase the nonlinear impairment tolerance [17-19], reduce the spectral footprint of single-channel signals [16], and therefore reduce both receiver complexity [20] and the required channel spacing in WDM systems [21, 22]. Of particular interest are sinc-shaped Nyquist pulses, which have a rectangular spectrum [16, 23, 24] and confine the signal to its Nyquist bandwidth [16]. This enables highest intra-channel spectral efficiencies, and has recently enabled transmission with a net spectral efficiency of 15 bit/s/Hz [25]. Combining a number of adjacent spectra, we end up with Nyquist WDM, where the carrier spacing is equal to the symbol rate (assuming identical symbol rates in all bands, which is not necessarily required). In contrast to CO-WDM where the phase of neighboring subcarriers is adjusted to minimize crosstalk [9], phase control of the carriers is not necessary for Nyquist WDM and OFDM [4]. Recently, this has been discussed as an option for Tbit/s super-channels [21], and favorable transmission properties have been predicted based on the fact that a train of modulated sinc-shaped Nyquist pulses has a relatively low peak-to-average power ratio [23]. Subsequently, several successful experiments demonstrated WDM with Nyquist pulse shaping and small guard bands [5, 26]. Also, a first demonstration of Nyquist WDM at 400 Gbit/s with a net spectral efficiency of 3.7 bit/s/Hz has been shown [27]. An arbitrary waveform generator (AWG) was used to create a Nyquist signal that was computed offline using a 601-tap finite impulse response (FIR) filter.

Nyquist WDM transmission with real-time sinc-pulse shaping and 16QAM has not yet been shown. The problem lies in the limited-length representation of acausal sinc-pulses in systems with real-time signal processing, where a practicable number of FIR filter taps has to be used. So far it had not been clear if a real-time computation will support Nyquist WDM transmission over significant distances.

In this paper, we report single-laser Nyquist WDM super-channel transmission at a record high aggregate line rate of 32.5 Tbit/s. This is the largest aggregate line rate ever encoded onto a single laser. In contrast to other experiments [5, 26, 27], we use neither offline processing at the transmitter nor guard bands. The electrical Nyquist signals are computed in real-time using a 64-tap FIR filter. The net spectral efficiency is 6.4 bit/s/Hz. We show transmission over 227 km. The achieved transmission distance is more than four times longer than what was reported for the most recent OFDM super-channel experiment [4]. The present experiment further shows how frequency comb generation is maturing. Here, 325 frequency locked carriers are generated from one source. This enables Tbit/s Nyquist WDM transmission with sufficient OSNR to reach distances of several hundred kilometers.

## II. BENEFITS AND CHALLENGES OF NYQUIST WDM TRANSMISSION SYSTEMS

There are significant advantages and challenges in the implementation of Nyquist WDM as compared to other schemes like standard WDM, OFDM and all-optical OFDM transmission systems. True Nyquist WDM builds on immediately neighbored partial spectra that do not overlap due to their rectangular shape. The bandwidth ($B_N$) of the partial spectra of channel $N$ is given by the Nyquist bandwidth of the encoded data, and it is equal to the symbol rate $R_N$. To achieve the rectangular partial spectrum, sinc-shaped Nyquist pulses are needed.

Due to the minimized bandwidth, Nyquist pulse transmitters have substantial benefits compared to standard non-return-to-zero transmitters (NRZ), because the available electrical bandwidth of components like digital-to-analog converters (DAC), driver amplifiers and modulators is optimally used. This, however, comes at the price of an increased amount of digital signal processing. Also, some oversampling is needed to accommodate anti-aliasing filters after the digital to analog converters [23].

A challenge when generating true Nyquist WDM is that the distance of the neighboring optical carriers has to be controlled precisely. For a true Nyquist WDM signal, the following condition has to be met: If two neighboring channels, namely channel $N$ and channel $N+1$, operate at symbol rates $R_N$ and $R_{N+1}$, respectively, the spacing $\Delta f$ has to be $\Delta f = (R_N + R_{N+1})/2$. If the spacing is larger, optical bandwidth is wasted. If the spacing is smaller, linear crosstalk from neighboring Nyquist channels will increase significantly. In our experiment we investigate the case, where all carriers have the same symbol rate $R$. In this case, the carrier spacing $\Delta f$ is equal to the symbol rate $R$.

Nyquist WDM transmission, when compared to all-optical OFDM [4], has the distinct advantage that only the dispersion within the bandwidth ($B_N$) of one Nyquist channel has to be compensated. Usually this is not done with dispersion compensating fibers, but rather by digital signal processing in the coherent receiver as it is the case for the present experiments. Electronic dispersion compensation can also be implemented for individual or small groups of subcarriers of an OFDM signal. Because the smallest optical filter bandwidth is limited by technical constraints, the all-optical OFDM symbol rate is in the same order of magnitude as the symbol rate of the Nyquist WDM channels. As the bandwidth of each modulated OFDM subcarrier is significantly larger than the OFDM symbol rate $R_N$, a larger amount of digital signal processing for dispersion compensation is required when compared to Nyquist WDM. Additionally, simulations in [23] indicate that Nyquist pulse shaped signals exhibit a lower peak to average power ratio (PAPR) compared to OFDM.

In Nyquist WDM receivers, the WDM channels are coarsely selected with optical filters. The final selection of a channel is done in the electrical domain by digital signal processing whereby sharp-edged filters can be realized. An additional challenge arises when implementing a clock recovery for Nyquist signals. As they will not have noticeable spectral components at the frequency of the sampling clock, most standard clock recovery algorithms will not work. Our solution for a clock recovery has been presented in [23].

## III. NYQUIST WDM SYSTEM CONCEPT

The envisioned Nyquist WDM super-channel concept with transmitter and receiver is illustrated in Fig. 1. The transmitter can be separated into two main parts: the carrier generation and the carrier modulation.

The carrier generation is a key part of this scheme as a small frequency drift of the optical carriers will immediately lead to an increased crosstalk from and to neighboring Nyquist WDM channels. It therefore makes sense to use an optical comb source where the generated optical carriers are inherently equidistant in frequency. This comb source could be replaced by 325 precisely-stabilized narrow line-width lasers. However, as the comb source only requires a single mode-locked laser, two EDFAs, a highly nonlinear fiber (HNLF) and a waveshaper (WS), it is doubtful that 325 lasers with additional frequency stabilization circuits could be operated with similar energy efficiency. To separate these carriers, we propose to use a cascade of an optical interleaver (IL) and two optical wavelength demultiplexers (DMX).

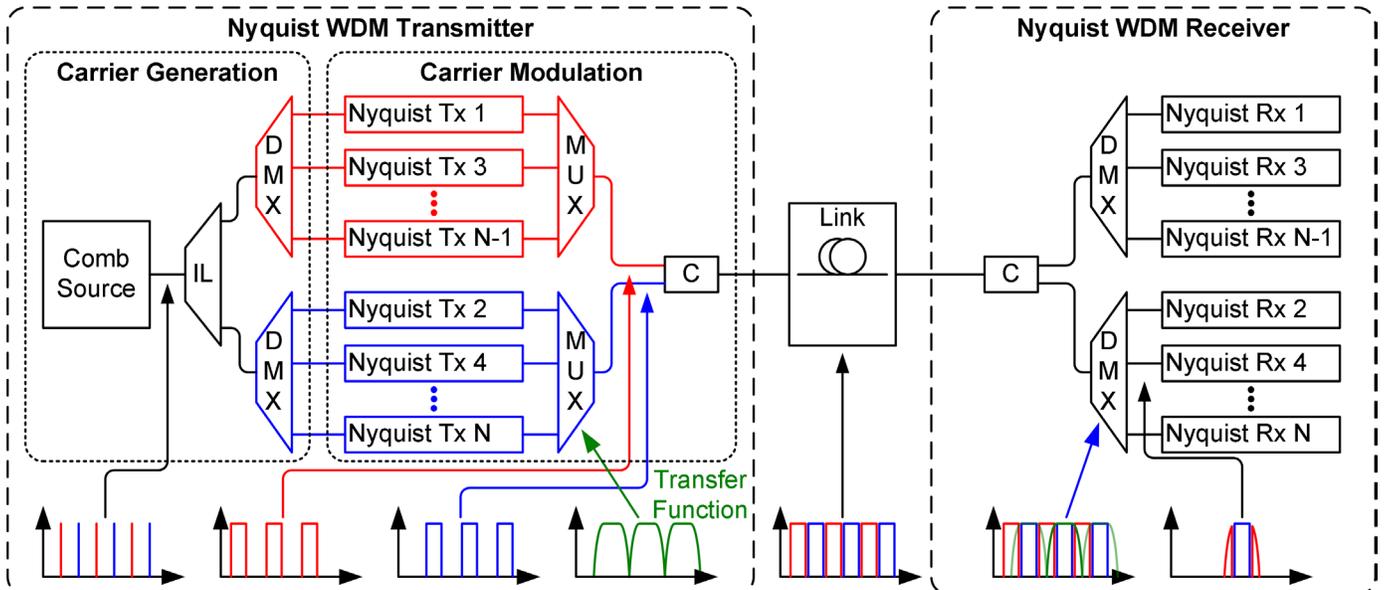

Fig. 1: Concept for a Nyquist WDM transmission system. In the transmitter, an optical comb source generates the optical carriers. Filters, namely an interleaver (IL) and an optical demultiplexer (DMX), separate the optical carriers. Nyquist transmitters (Nyquist TX1, 2, … *N*) encode the data. Multiplexers (MUX) and a standard optical coupler (C) combine the transmitter outputs. After transmission, a coupler (C) and optical demultiplexers (DMX) split the carriers for the Nyquist receivers (Nyquist RX1, 2, … *N*). The schematic spectra illustrate the Nyquist WDM transmitter and receiver concept for a total number of *N* = 6 optical carriers. Spectra of odd carriers (———), spectra of even carriers (———), and MUX/DMX transfer functions (———) are shown.

For carrier modulation, we propose the use of Nyquist transmitters with digital signal processing (DSP), as these transmitters can generate signal spectra with an extremely steep roll off [23, 24]. To maintain the rectangular shape of the Nyquist spectra, special care has to be taken when combining different Nyquist channels. To this end, we select a combination of two optical multiplexers and an optical coupler. The two optical multiplexers combine odd or even channels, respectively, without limiting the Nyquist spectra. The optical coupler combines odd and even channels for transmission over the optical fiber link. The transmitter scheme is depicted on the left hand side of Fig. 1.

In the Nyquist WDM receiver, the signal is split by a coupler, and two optical demultiplexers separate odd or even optical carriers in front of the Nyquist receivers. Even though it would be possible to simply split the signal and receive it using multiple coherent Nyquist receivers, we suggest including optical demultiplexers for two reasons: First, this reduces the insertion loss of the receiver structure significantly. Second, as the total optical power of the signal in a coherent receiver is restricted by physical limitations of the photodiodes, the power of the received signal can be increased significantly, and unnecessary power loading of the balanced detectors with a large number of unwanted carriers is avoided. After filtering, residual components of neighboring Nyquist channels remain (see bottom right inset in Fig. 1). These residual spectra have to be removed by digital brick wall filtering in the Nyquist receivers.

IV. IMPLEMENTATION OF A NYQUIST PULSE TRANSMITTER

Pulse shaping is crucial for the implementation of Nyquist WDM transmission systems. Sinc-shaped Nyquist pulses extend infinitely in time and generate a rectangular spectrum (insets in Fig. 2). The Nyquist pulses repeat with the impulse spacing $T$, are modulated with complex data, and have a total bandwidth $B = 1/T$, which equals the symbol rate $R$. For our experiment, two real-time Nyquist pulse transmitters are implemented (one is shown in Fig. 2) to modulate odd and even carriers. The transmitter setup is based on the setup in [23, 24], which was developed from our multiformat transmitter [28]. For an efficient pulse shaping, we use oversampling with two samples per symbol. In a first step, a pseudo-random bit sequence (PRBS, length $2^{15}-1$) is generated in real-time by the two FPGAs (Xilinx XCV5FX200T). The 16QAM symbols enter a FIR filter with 64 taps, thereby generating sinc-shaped Nyquist pulses modulated with the data. The number of taps directly impacts the processing delay of the transmitter and is limited by the available space for logic on the FPGA. The resulting digital signal is then converted to the analog domain using either two VEGA DAC25 (Tx1 in Fig. 3 (a)), or two VEGA DAC-II (Tx2 in Fig. 3 (a)). We use different DACs for the transmitters due to their availability in our laboratory. We did not observe a significant performance difference of the two transmitters. Simulated eye diagrams are displayed as insets in Fig. 2. The DACs operate at 25 GSa/s, generating 12.5 GBd signals with an electrical bandwidth of 6.25 GHz. In contrast to [23, 24], we use electrical lowpass filters with a 3 dB bandwidth of 12 GHz and a suppression of > 30 dB at 13 GHz to remove the image spectra. After amplification,

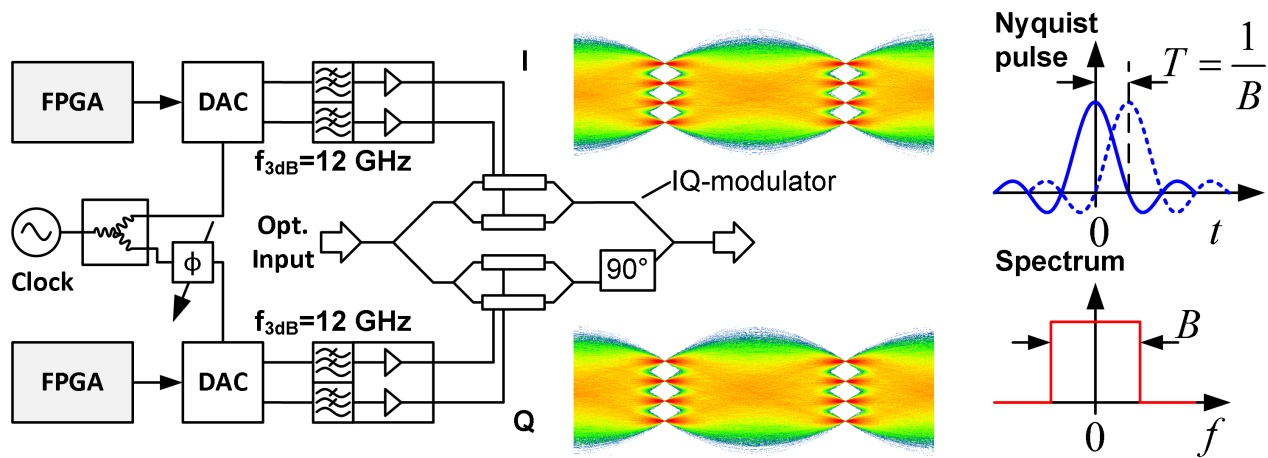

Fig. 2: Implementation of real-time transmitter for sinc-shaped Nyquist pulses. Two FPGAs are programmed to generate the PRBS ($2^{15}$-1), to perform symbol mapping, and to provide pulse shaping with an FIR-filter. The in-phase (I) and quadrature (Q) signals are converted into the analog domain using two high speed VEGA-DACs. Image spectra are removed with additional anti-aliasing filters, and the resulting signals are amplified for driving the optical IQ-modulator. Insets: Simulated eye diagrams for I and Q drive signals, sinc-pulse and corresponding spectrum at the output of the transmitter. The simulated eye diagrams illustrate that there is only a small tolerance in clock phase when receiving such a signal, which makes a proper clock phase recovery procedure extremely important [23].

the signals are fed to an optical IQ-modulator which in turn modulates the optical carrier [23].

## V. NYQUIST WDM EXPERIMENT

Our Nyquist WDM system Fig. 3 (a)–(c) consists of four main components: the optical comb source, two Nyquist WDM transmitters as in Fig. 2, a polarization multiplexing emulator, and a coherent optical Nyquist WDM receiver.

The optical comb source is actually one of the key components in this experiment. It did not only provide a cost-effective and energy-efficient way to generate a large number of optical carriers, it also generated them with a highly stable frequency spacing, which is useful for the case of equal symbol rates in all channels having a carrier spacing equal to the symbol rate. This comb source uses a pulse train from an ERGO-XG mode-locked laser (MLL) which is amplified and filtered to remove amplified spontaneous emission. The MLL output is split in two parts, one of which is spectrally broadened in a highly nonlinear photonic crystal fiber [4]. In the waveshaper (WS), the original MLL comb is bandpass-filtered and fills the void in the notch-filtered broadened comb such that unstable sections in the center of the broadened spectrum are replaced by the original MLL spectrum. This spectral composing process is also exploited for equalizing the frequency comb to form a flat output spectrum. For Nyquist WDM transmission there is no requirement for a stabilization scheme to guarantee a fixed initial phase of all carriers relative to the beginning of a symbol time slot (as is the case for coherent WDM [9]). A number of 325 optical carriers are generated between 1533.47 and 1566.22 nm with a spacing of 12.5 GHz. Our measurements indicate that the line width of the carriers is significantly lower than the line width of our local oscillator in the receiver (Agilent 81682A – external cavity laser – ECL – line width typ. 100 kHz). The MLL is adjusted such that the carriers fall on the ITU grid. This allows us to use off-the-shelf optical components.

For modulation, the spectral lines are decomposed into odd and even carriers using a standard optical interleaver. Odd and even carriers are modulated with Nyquist transmitters Tx1 and Tx2, respectively, see Fig. 3. To generate a Nyquist WDM signal, the symbol rate of 12.5 GBd is chosen to equal the carrier spacing of 12.5 GHz. Both transmitters operate with separate sampling clock sources as no symbol synchronization is required. After modulation, odd and even carriers are combined in an optical coupler to form the Nyquist WDM signal. The two outputs of this coupler are then delayed relative to each other for data de-correlation (delay 5.3 ns) and combined in a polarization beam combiner to emulate polarization multiplexing.

The signal is then amplified, and transmitted over distances of 75.78 km and 227.34 km using a Corning SMF-28 with EDFA-only amplification. The optimum launch power was found to be 18 dBm for the complete Nyquist WDM signal. This corresponds to a power of –7 dBm per carrier. This launch power was optimized for the carrier at 1550.015 nm.

After transmission, the carrier of interest is selected in a WS, amplified, and finally received in an optical modulation analyzer (OMA – Agilent N4391A). Four signal processing steps were performed before the 16QAM demodulation. First, the chromatic dispersion is compensated. Second, a digital brick wall filter selects one channel and removes all remainders of neighboring channels. Third, the standard polarization tracking algorithm [29] separates the two polarizations, and fourth, the clock phase is recovered as described in [23]. Only the digital brick wall filtering and the clock phase estimation algorithm had to be implemented in addition to the standard algorithms included in the OMA software. No modification of the 16QAM receiver algorithms of the OMA software was required. Using the built-in frequency offset estimation algorithm, we were able to precisely determine the frequency offset up to 500 MHz. This allows for frequency tracking without requiring pilot symbols, which would lead to additional overhead. A least mean square adapted linear FIR filter with 51 taps was used as equalizer to compensate for the frequency dependence of the overall transmission system. PMD was not compensated for.

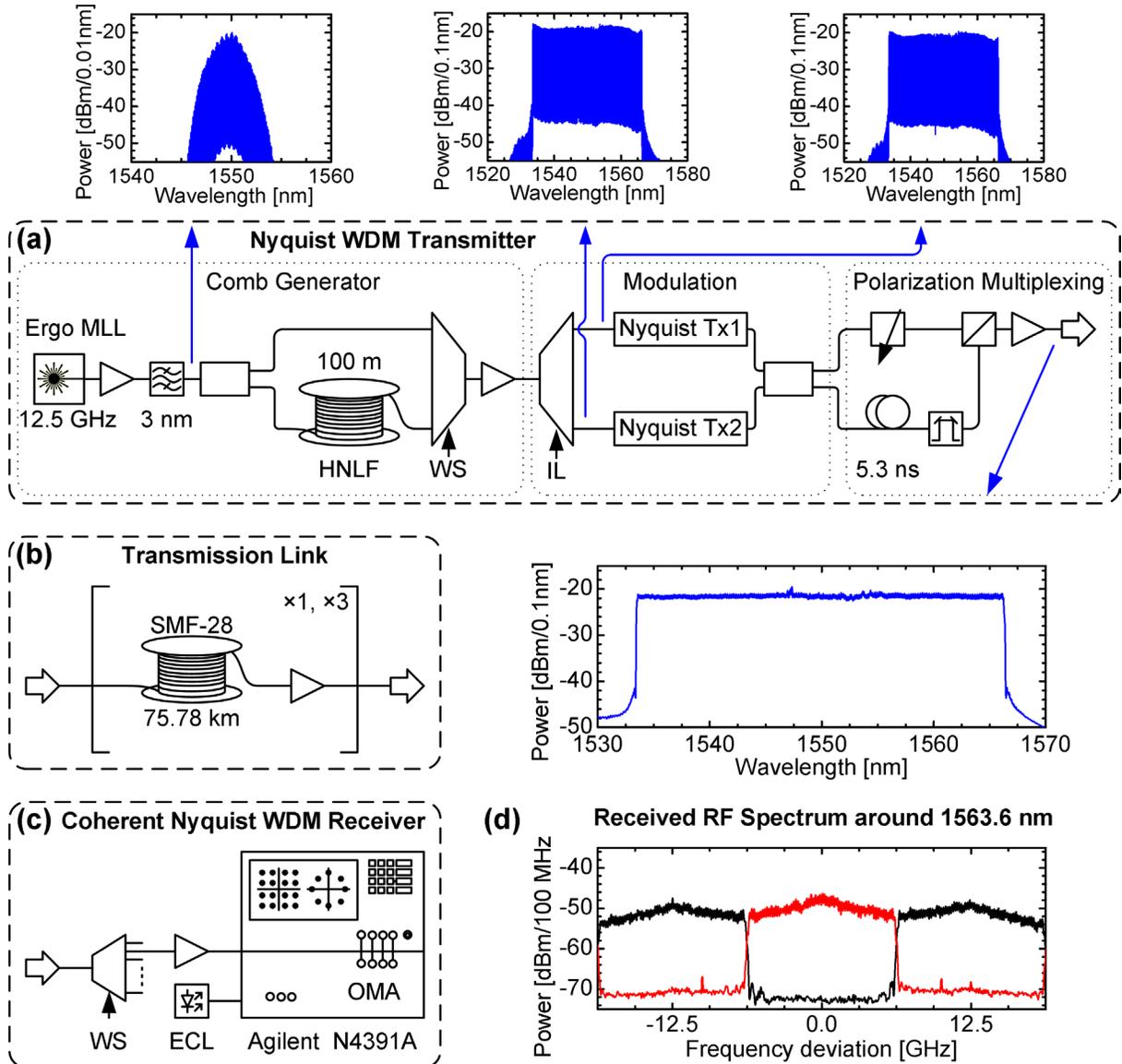

Fig. 3. Nyquist WDM setup. (a) A mode-locked laser (MLL) produces a frequency comb (see inset above (a)) which is broadened in a highly nonlinear fiber (HNLF). A waveshaper (WS) equalizes the resulting comb of 325 lines and replaces the unstable central part by a copy of the original MLL comb. An optical interleaver (IL) separates odd and even carriers (power spectra in the inset above (a)). On these carriers the transmitters Tx1 and Tx2 (see schematic in Fig. 2) encode 16QAM data in form of sinc-shaped Nyquist pulses. Polarization multiplexing is emulated. The arrow points to the resulting optical power spectrum (inset below (a)). (b) The optical signal is then transmitted over one or up to three spans of Corning SMF-28 with EDFA-only amplification. (c) In a coherent Nyquist WDM receiver a WS selects a 60 GHz wide group of Nyquist channels fitting to the bandwidth of the optical modulation analyzer (OMA). An external cavity laser (ECL – Agilent 81682A) provides the local oscillator for the coherent receiver. (d) Two-sided RF power spectrum after down conversion from an optical carrier at 1563.6 nm (——) and from the adjacent carriers (——). The rectangular shape of the spectra proves the effectiveness of the Nyquist pulse shaping.

To characterize the signal quality, we measured the error vector magnitude (EVM). We derived a bit error ratio (BER) estimate from the EVM data [30]. We verified this estimate for selected points to support the applicability of this estimation technique. The BER was measured with the OMA for carriers in the back-to-back case and for the transmission over 227.34 km. We chose carriers that exhibited higher EVM values, which enabled us to provide an upper limit of the BER. The accuracy of the BER estimation method from [30] has been demonstrated experimentally in [31] and was again confirmed in these experiments.

## VI. EXPERIMENTAL RESULTS

Back-to-back measurements serve as a reference for the overall system performance, Fig. 4 (a). The EVM for almost all carriers was below the threshold for second generation FEC (BER = $2.3 \times 10^{-3}$). The total EVM (defined as the root mean square of the EVM of all carriers) was $EVM_{tot} = 10.3$ %. In Fig. 4 (b, c) we show the results after transmission. $EVM_{tot}$ degrades by 1.0 percentage points for a distance of 75.78 km and by 1.7 percentage points for a distance of 227.34 km. The EVM for all carriers and distances is well below the limiting EVM for a BER of $1.8 \times 10^{-2}$ using next generation soft decision FEC [32]. The EVM differences between the two polarizations, especially in the outer wavelength range, are due to the wavelength dependence of the 3 dB coupler in the polarization multiplexing scheme. An additional degradation can be traced back to the non-ideally gain-flattened EDFAs that were available for the experiment. These EDFAs lead to the uneven received spectra after transmission. If EDFAs with better gain flattening were used in such an experiment, we would expect a significant increase in achievable transmission distance.

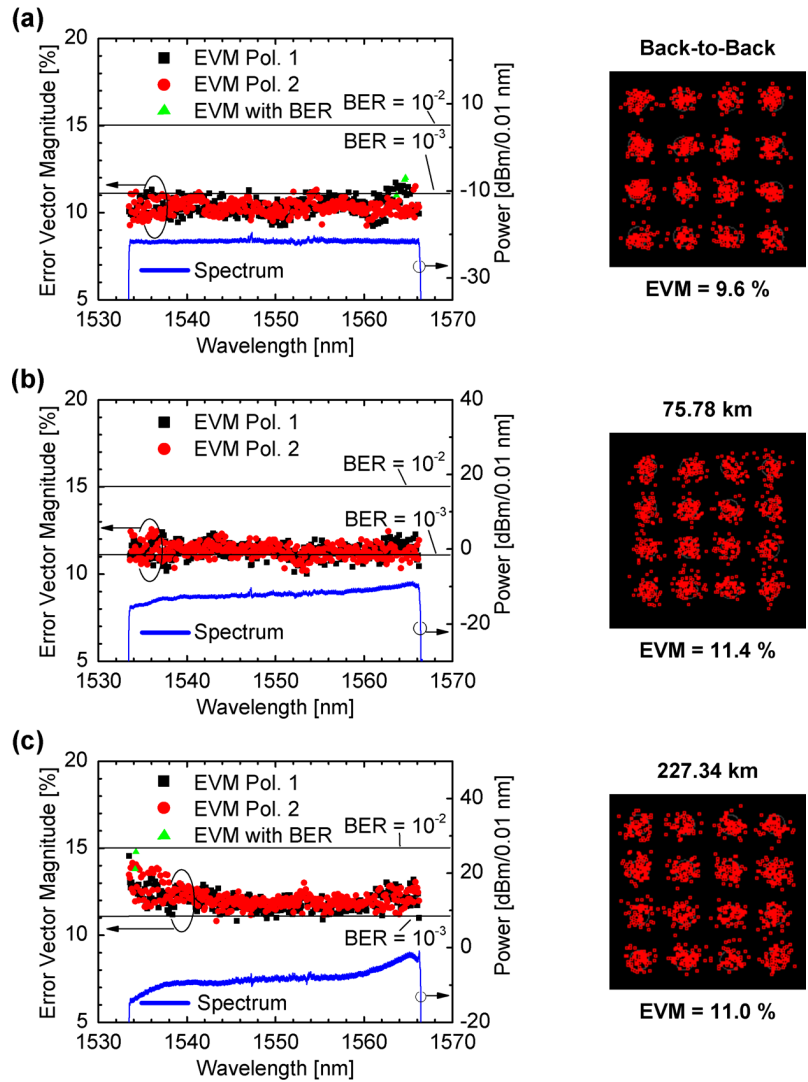

Fig. 4: Experimental results for the transmission experiment. The measured error vector magnitudes for all subcarriers (■ – polarization 1, ● – polarization 2) and the received optical spectra (——) are plotted for (a) back-to-back characterization, (b) transmission over 75.78 km, and (c) transmission over 227.34 km. For selected EVM values in the area around the highest EVM values (▲), the BER was measured and the results are displayed in Table I to verify the applicability of BER estimation from EVM measurements. Corresponding constellation diagrams for the carrier at 1550.015 nm are shown for the various transmission distances.

When approaching the Nyquist spacing of the channels, a certain amount of linear crosstalk is to be expected due to the finite filter slopes. The finite impulse response (FIR) filters are implemented by digital real-time signal processing [23, 24]. In the present experiment, the crosstalk is very small as shown in the RF spectra in Fig. 3 (d). Here we show the measured RF spectra when only transmitter 1 (——) or transmitter 2 (——) are turned on. A small amount of sampling clock leakage in the DAC can be observed at 12.5 GHz. The tones around ±10 GHz originate from the sampling oscilloscope in the receiver. The EVM degradation due to residual crosstalk is measured to be 1.5 to 2 percentage points. To quantify the influence of linear crosstalk from neighboring Nyquist channels, we measured the BER for the wavelengths 1563.66 nm and 1564.58 nm with and without neighboring channels. For 1563.66 nm, the BER increased from $6.3 \times 10^{-4}$ without neighboring channels to $1.2 \times 10^{-3}$ with neighboring channels. For 1564.58 nm the BER increased from $1.4 \times 10^{-4}$ to $2.1 \times 10^{-3}$. This shows that the back-to-back performance is mainly limited by linear crosstalk due to the limited number of taps. This crosstalk could be reduced by increasing the carrier spacing from the Nyquist case to a larger spacing. This could increase the achievable transmission distance or reduce the required amount of FEC overhead — however, such a system would no longer be a Nyquist WDM system. The line rate of 32.5 Tbit/s corresponds to a net data rate of 26 Tbit/s for the transmitted signals (taking into account the 25 % FEC overhead [32]).

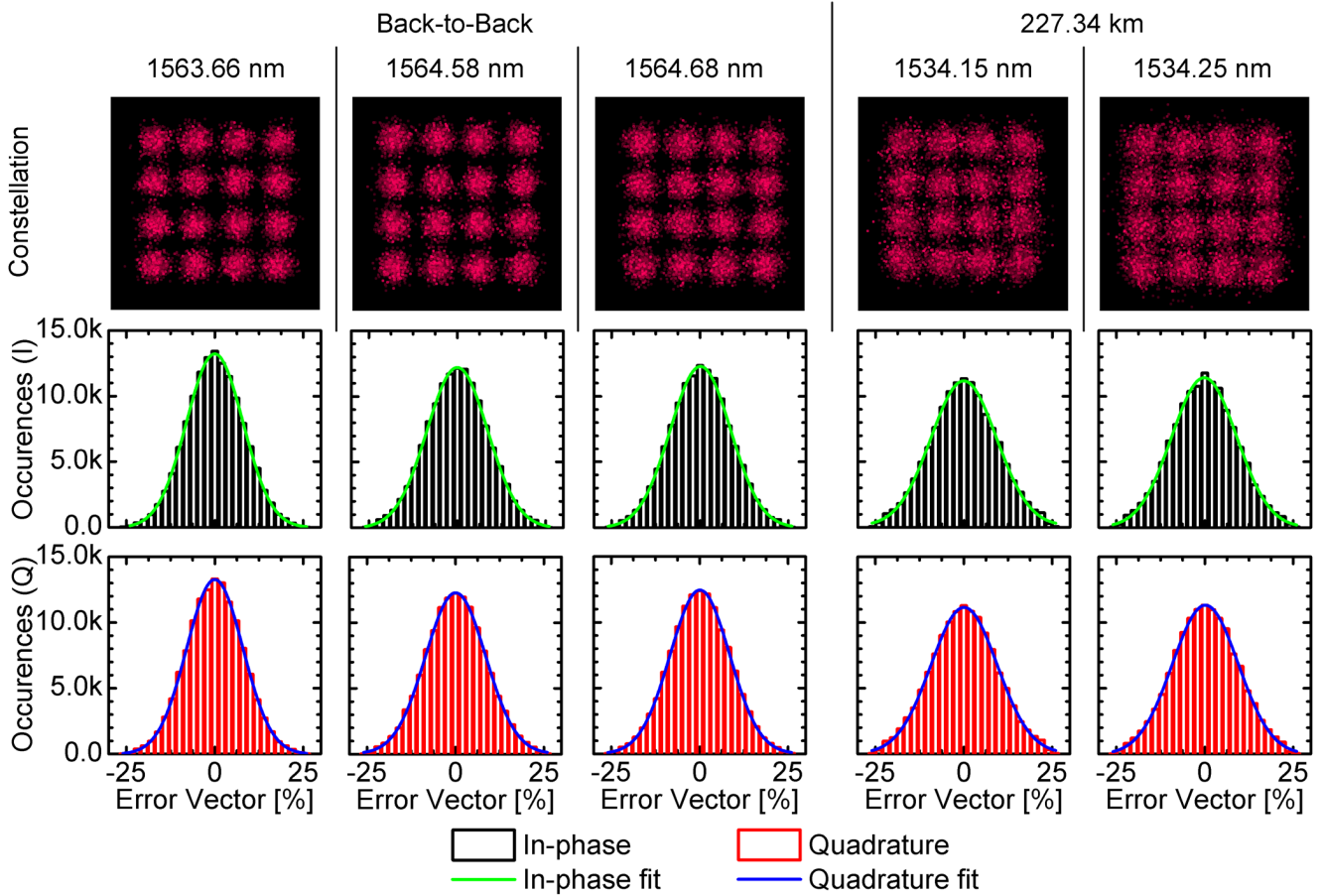

Fig. 5: Statistical analysis of selected carriers. We show constellation diagrams of the carriers used for checking the validity of the assumptions for the BER-EVM relationship presented in Eq. (1). We performed a statistical analysis of the in-phase (I ———) and quadrature-phase (Q ———) error vectors. The respective Gaussian fits (for I —— and Q ——) indicate a Gaussian probability density function of the added noise as required for the reported BER estimations.

To verify the relationship between EVM and BER, we chose to measure BER and EVM for some selected carriers. Due to time constraints, we measured the BER only for the carriers presented in Table I. We chose these carriers such that the worst-case EVM for the back-to-back and 227 km transmission was included. In Table I we show the measured EVM values and the measured BER. In addition, we calculated a BER corresponding to the measured EVM values using Eq. (1), which was derived in reference [30]:

$$\mathrm{BER} \approx \frac{1-M^{-\frac{1}{2}}}{\frac{1}{2}\log_2 M} \mathrm{erfc}\left[\sqrt{\frac{3/2}{(M-1)(k\,\mathrm{EVM}_m)^2}}\right]. \quad (1)$$

For our 16QAM signal the number of all possible constellation points is $M = 16$, the number of bits encoded in one QAM symbol is $\log_2 M = 4$, and the modulation format dependent factor is $k^2 = 9/5$. It has to be mentioned that a misprint happened in equation (4) in reference [31], where $\sqrt{2}$ has to be replaced by 1. However, the corresponding plots in Fig. 3b of reference [31] were calculated correctly. The calculated BER differs only slightly from the measured BER, supporting the applicability of the BER-EVM relationship, which is based on a Gaussian hypothesis.

The distribution of the noise on the constellation points is a critical factor for the applicability of the BER-EVM relationship, and for predicting error-free operation when using soft-decision FEC [32]. To support the claim that our constellation field vectors are perturbed by additive Gaussian noise, we plotted in Fig. 5 the constellation diagrams for the carriers listed in Table I. Additionally, we performed a statistical analysis of the distribution of in-phase (I) and quadrature-phase (Q) error vectors in the constellation diagram. The results are displayed in Fig. 5 and support the claim of a Gaussian distribution of the added noise.

In summary, our results show that a FIR filter with 64 taps suffices for implementing a Nyquist WDM transmis-sion system when using twofold oversampling. However, we expect an additional performance improvement when longer filter lengths are used. The quality of reception could not be compared to 16QAM WDM experiments as linear crosstalk from neighboring carriers limited the EVM. Also, as all carriers in this experiment were derived from a single laser, the experiment should not be compared to experiments using a large number of separate lasers. Here, we were able to increase the data rate by 20 % and the transmission distance by a factor of 4.5 compared to our previous experiment [4] with record-high data rates transmitted on a single laser.

TABLE I
COMPARISON OF ESTIMATED AND MEASURED BER. WE MEASURED BER AND EVM FOR THE CARRIERS PRESENTED BELOW AND CALCULATED AN EQUIVALENT BER TO VERIFY THE EVM – BER RELATIONSHIP.

| Distance | Wavelength [nm] | EVM [%] | Calculated BER | Measured BER |
|---|---|---|---|---|
| B2B | 1563.66 | 11.0 | $1.2 \times 10^{-3}$ | $1.2 \times 10^{-3}$ |
|  | 1564.58 | 11.9 | $1.9 \times 10^{-3}$ | $2.1 \times 10^{-3}$ |
|  | 1564.68 | 12.0 | $2.0 \times 10^{-3}$ | $2.3 \times 10^{-3}$ |
| 227 km | 1534.15 | 13.8 | $5.9 \times 10^{-3}$ | $5.7 \times 10^{-3}$ |
|  | 1534.25 | 14.8 | $9.1 \times 10^{-3}$ | $9.4 \times 10^{-3}$ |

## VII. CONCLUSION

In this paper we show that Nyquist WDM is a promising candidate for next generation communication systems. Nyquist WDM improves spectral efficiency and transmission distance when compared to previously investigated all-optical OFDM systems. We demonstrate for the first time 16QAM Nyquist WDM transmission with a symbol rate equal to the carrier spacing. The sinc-pulse shaping is done by real-time digital signal processing. A total aggregate data rate of 32.5 Tbit/s and a net spectral efficiency of 6.4 bit/s/Hz are achieved. As all carriers are generated from a single laser, this is a new data rate record when using a single laser source.

## ACKNOWLEDGEMENT

The authors acknowledge partial funding from the Karlsruhe School of Optics and Photonics (KSOP), and from the German Research Foundation (DFG). We thank Sander Jansen of Nokia Siemens Networks and Beril Inan of Technical University of Munich for lending equipment and for discussions. Support by the Xilinx University Program (XUP), the Agilent University Relations Program, the European network of excellence *EuroFOS* and the European research project ACCORDANCE is gratefully acknowledged.